\documentclass[12pt,onecolumn]{article}

\usepackage{arxiv}
\usepackage{amsmath}
\usepackage{graphics}
\usepackage[utf8]{inputenc} 
\usepackage[T1]{fontenc}    
\usepackage{url}            
\usepackage{booktabs}       
\usepackage{amsfonts}       
\usepackage{nicefrac}       
\usepackage{microtype}      
\usepackage{lipsum}

\usepackage{amsfonts}
\usepackage{amssymb}
\usepackage{bbold}
\usepackage{booktabs}
\usepackage{ulem}
\usepackage{xr}
\usepackage[justification=centering ]{subcaption}
\usepackage{graphicx}
\usepackage{amsmath}
\usepackage[usenames,dvipsnames]{xcolor}
\usepackage{multirow}
\usepackage{lscape}
\usepackage[sorting = none, backend = bibtex, style=phys]{biblatex}
\usepackage{amsfonts}
\usepackage{amssymb}
\usepackage{bbold}
\usepackage{booktabs}
\usepackage{ulem}
\usepackage{xr}
\usepackage[justification=centering ]{subcaption}
\usepackage{tikz}
\usetikzlibrary{shapes}
\usepackage{soul}
\usepackage{nicefrac}
\usepackage{xcolor}
\usepackage{bm}
\usepackage{multirow}
\usepackage{amsfonts}
\usepackage{amssymb}
\usepackage{bbold}
\usepackage{hyperref}
\usepackage{capt-of}
\usepackage{float}
\usepackage{lipsum}
\usepackage{soul}
\usepackage{color}

\newcommand{\Ntracers}{N}

\renewcommand{\vec}[1]{\boldsymbol{\bm{#1}}}

\title{Lagrangian analysis of turbulent blood flow in the human left heart}
\author{
Fabio Guglietta\\
  Department of Physics \& INFN, University of Rome ``Tor Vergata'', Rome, Italy.\\
  \texttt{fabio.guglietta@roma2.infn.it} \\
\And
  Martino Andrea Scarpolini\\ 
  Gran Sasso Science Institute (GSSI), L'Aquila, Italy, \& INFN Tor Vergata, Rome, Italy.\\
\And
Francesco Viola\\ 
  Gran Sasso Science Institute (GSSI), L'Aquila, Italy.\\
\And
Luca Biferale\\
  Department of Physics \& INFN, University of Rome ``Tor Vergata'', Rome, Italy.\\}

\begin{document}
\maketitle
\begin{abstract}
We present a Lagrangian analysis of turbulent blood flow in the human left heart using high-fidelity simulations based on a patient-specific anatomical model. Leveraging a fully coupled fluid–structure–electrophysiology interaction (FSEI) framework, we track the motion of Lagrangian (passive) tracers to investigate the multiscale statistical properties of velocity fluctuations {over more than four decades}. Our analysis reveals strong Lagrangian intermittency throughout the left heart, reflecting the complex and unsteady nature of cardiovascular flow.
The present work underscores the sensitivity of Lagrangian statistics to physiological parameters and highlights their potential for improving the understanding of pathological flow conditions in cardiovascular systems.
Such Lagrangian tool provides a statistical foundation for modelling shear-induced damage in red blood cells (hemolysis), with implications for the evaluation of prosthetic valves and blood-contacting medical devices.
\end{abstract}

\maketitle
\setlength{\parskip}{0pt}
\section{\label{sec:intro}Introduction}
Numerical simulations of the human heart provide a powerful tool to investigate physical and statistical properties of the blood flow~\cite{verzicco_2022_EFM,santiago2018whole_heart_fsi,bucelli_2023_efsi,davey2024simulating}. Understanding the dynamics of turbulent blood flow in the cardiovascular system is essential for advancing our knowledge of hemodynamics and its implications in both physiological and pathological conditions. In particular, the left heart presents a uniquely complex flow environment due to pulsatile forcing, deformable boundaries, and strong interactions with valve motion.

While many works have focused on cardiovascular hemodynamics, relatively few studies have examined turbulent features from a rigorous statistical perspective. Some image-based computational studies have analyzed blood flow in the aorta~\cite{cheng2025characteristics,lantz2012quantifying} and left heart~\cite{chnafa2014image,chnafa2016image,bennati2023turbulent}, with particular attention to the onset of turbulence and its implications for hemolysis and thrombus formation~\cite{bennati2023turbulent}. Nitti et al.~\cite{nittiNumericalInvestigationTurbulent2022} conducted a local Eulerian analysis of turbulence correlations and energy spectra in mechanical aortic valves using a finite-difference flow solver with a direct-forcing immersed boundary method. These works typically rely on Eulerian diagnostics, such as turbulent kinetic energy, to assess the degree and evolution of turbulence.

However, Eulerian approaches are inherently limited in the context of the heart, where large geometric deformations can make fixed-point measurements complex or even impossible. For example, a probe located near the ventricular apex during diastole may fall outside the chamber during systole. This makes collecting consistent data throughout the cardiac cycle difficult and often leads to biased or incomplete turbulence characterization. In contrast, a Lagrangian perspective, where fluid elements are tracked along their trajectories, naturally adapts to the evolving domain. Passive tracers can follow the flow even in highly deformable regions, providing an unbiased and continuous view of the velocity field throughout the heart.

Lagrangian methods thus offer unique advantages in cardiovascular applications, not only by accommodating moving boundaries but also by enabling multiscale statistical analyses~\cite{toschiLagrangianPropertiesParticles2009}. From tracer trajectories, one can extract detailed information about velocity increments, acceleration statistics, and temporal correlations~\cite{toschiLagrangianPropertiesParticles2009,biferale2008lagrangian,Benzi2023,bentkamp2019persistent}. These tools are particularly well suited for confined and inhomogeneous domains like the heart, where spatial variability and localized structures play a central role in characterizing flow dynamics. Moreover, such high-resolution flow data combined with Lagrangian tracers can be used to investigate the dynamics at the scale of red blood cells (RBCs) to assess, for example, shear-induced damage, which is key to understanding hemolysis and thrombosis in medical devices and pathological conditions~\cite{scarpolini2025hemodynamiceffectsintrasupra}. Indeed, the velocity gradients probed by the tracers can be used as a boundary condition for ab-initio numerical simulations at the scale of the RBC in order to simulate the dynamics of a single cell subjected to a realistic velocity field~\cite{guglietta2020effects,guglietta2021loading,guglietta2020lattice,Taglienti2024,biferale2014deformation}. 

In the context of patient-specific simulations, these tools can help the development of cardiac digital twins, which aim to combine anatomical imaging with physics-based modeling to simulate patient-specific cardiovascular function~\cite{viola_2023_digital_twins}. By incorporating Lagrangian turbulence diagnostics into digital twins, we can achieve more accurate assessments of flow-related risks, such as regions prone to shear stress accumulation or inefficient transport.

\begin{figure}[t!]
    \centering
    \includegraphics[width=1.\linewidth]{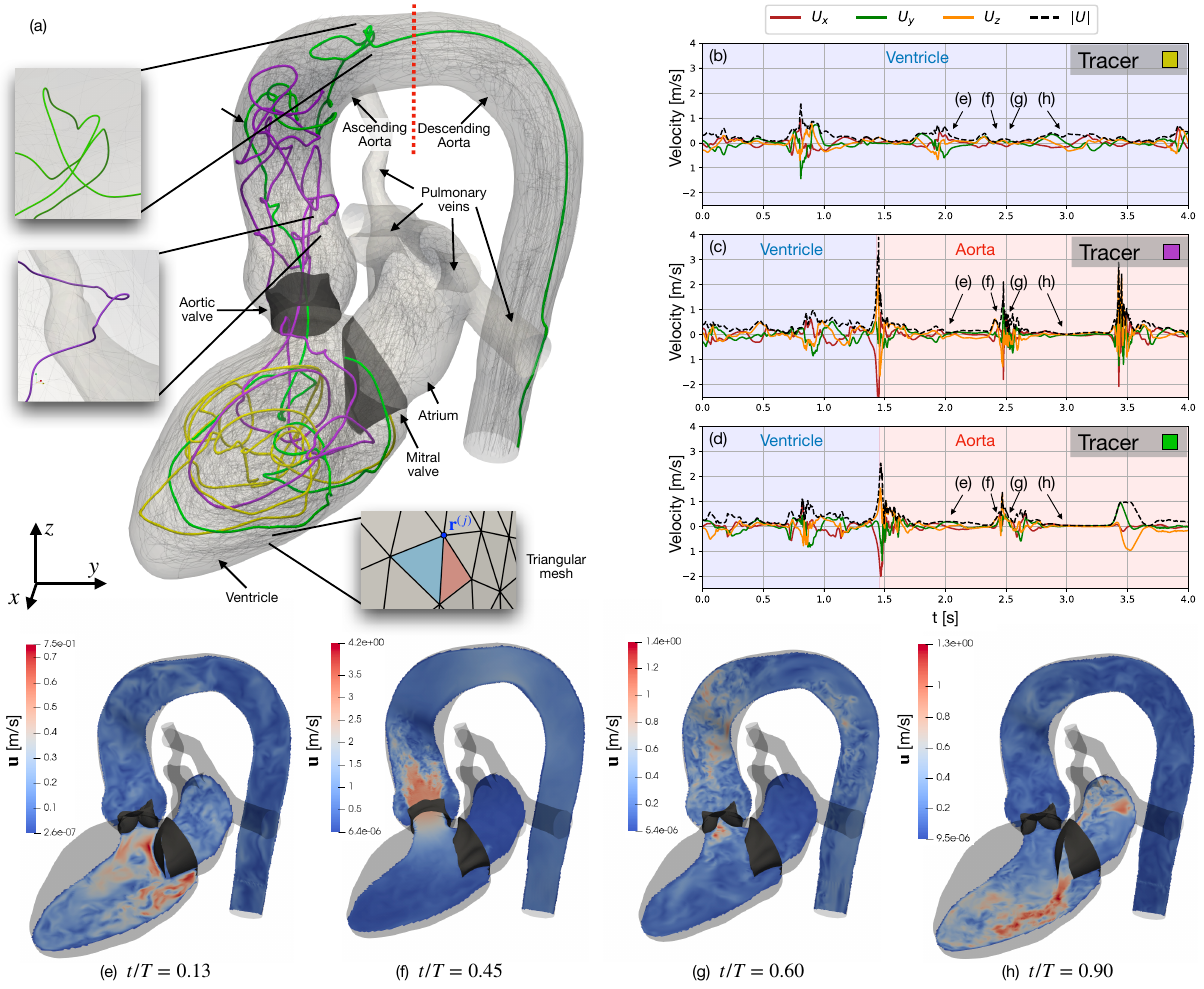}
    \caption{
    Panel (a) shows the 3D geometry of the left heart (the mitral and aortic valves are shown in darker color) with approximately one hundred tracer paths (gray lines), highlighting three examples in yellow, purple, and green. Insets on the left provide zoomed views of portions of the purple and green trajectories. The inset on the bottom provides a zoom of the triangular mesh used to model the heart: two neighboring triangles are highlighted in blue and red, while a node mesh with coordinates $\mathbf{r}^{(j)}$ is shown with a blue circle. {The red dashed vertical line represents the boundary used to divide the aorta into its ascending and descending parts, allowing us to investigate the two sections separately.}
    Panels (b-d) display the three velocity components ($U_x$, $U_y$, $U_z$) and the velocity magnitude $|\mathbf{U}|$ for each selected tracer over four cardiac cycles. The background shading indicates the chamber occupied by the tracer at each instant: blue for the ventricle and red for the aorta.
    {Panels (e–h) display the instantaneous fluid velocity (magnitude) $|\mathbf{u}|$ at selected phases of the cardiac cycle: end of diastole ($t/T = 0.13$), peak systolic ejection ($t/T = 0.45$), end of systole ($t/T = 0.60$), and beginning of diastole ($t/T = 0.90$). These phases are also reported in panels (b-d).}
    }\label{fig:sketch}
\end{figure}

In this work, we present a systematic Lagrangian analysis of turbulence in the left heart using high-fidelity simulations based on a patient-specific anatomical model [see Fig.~\ref{fig:sketch}, panel (a)]. We employ a fully coupled fluid–structure–electrophysiology interaction (FSEI) solver~\cite{viola2023fsei,viola2022fsei_gpu} to simulate the whole cardiac cycle, and we track hundreds of thousands of passive tracers to explore turbulence across a wide range of time scales, from one heartbeat to fractions of the viscous time-scale. Furthermore, 
our analysis focuses on the impact of two key physiological parameters: heart rate (HR) and aortic valve stiffness. By comparing simulations at HR = 40, 60, and 80 bpm, we quantify the effect of cardiac frequency on velocity fluctuations. Additionally, we investigate a pathological condition involving a stiffened aortic valve, mimicking  {aortic} stenosis~\cite{baumgartner2017ESC_AVS_assessment_echocardiography}, by increasing the leaflet stiffness.

To characterize turbulence, we employ a set of Lagrangian statistical tools, including time correlation functions, structure functions, local scaling exponents, and flatness factors. Correlation functions reveal temporal coherence and flow memory, while structure functions and their local slopes provide insight into the  {scale-by-scale} behavior of velocity fluctuations. Flatness, in turn, quantifies intermittency and the prevalence of rare, intense flow events. Together, these metrics allow for a comprehensive evaluation of turbulent dynamics under physiologically realistic conditions.

The paper is organized as follows: in Sec.~\ref{sec:numerical}, we first describe the numerical method used to perform simulations, and then we introduce some quantities typical of Lagrangian statistics. In Sec.~\ref{sec:results}, we show the results first at fixed heart rate and aortic valve elasticity (Sec.~\ref{sec:results-chamb}), and then at varying heart rate and valve elasticity (Sec.~\ref{sec:results-HR} and Sec.~\ref{sec:results-ke}, respectively). Conclusions are provided in Sec.~\ref{sec:conclusions}.


\section{\label{sec:numerical}Setup and numerical method}
In this work, we simulate the dynamics of $\Ntracers$ Lagrangian tracers in the human left heart (see Fig.~\ref{fig:sketch}). 
The triangular mesh used in this study to represent its endocardium is based on the high-fidelity anatomical model developed in Ref.~\cite{scarpolini2025hemodynamiceffectsintrasupra}, which was derived from a thoracic CT scan of a healthy 44-year-old male. The imaging was acquired using a 320-detector row CT scanner (Aquilion One PRISM edition, Canon Medical Systems) during the diastolic phase, achieving a spatial resolution of $0.728\times0.728\times1.0~\mbox{mm}$. 
The left heart model, which includes the left atrium, left ventricle, and aorta, was reconstructed using semi-automatic segmentation in 3DSlicer~\cite{3dslicer}. The atrium features both the auricle (left atrial appendage) and the pulmonary veins, which act as inlets for incoming blood. The mitral valve geometry was adopted from a previous numerical study~\cite{meschini2019mitral_stenosis}. 
The segmented heart corresponds to a left atrial volume of approximately $80~\mbox{ml}$ and a left ventricular volume of $114~\mbox{ml}$. The aortic portion includes the ascending aorta, the aortic arch, and the initial tract of the descending aorta, thus enabling a realistic representation of the blood outflow from the ventricle.

All simulations are performed using GPU acceleration (the code is written in CUDA Fortran~\cite{viola2022fsei_gpu}) to ensure computational efficiency, enabling the simulation of detailed cardiovascular dynamics within clinically acceptable time frames.
Simulations ran on two NVIDIA A100 80 GB GPUs. For the HR = 60~bpm case, each cardiac cycle lasts about 30 hours.

\subsection{Fluid-structure-electrophysiology interaction (FSEI)}
The computational model employed in this work couples fluid dynamics, structural mechanics, and electrophysiology in a fully coupled fluid–structure–electrophysiology interaction (FSEI). 
This framework has already been largely used and validated in previous works~\cite{viola2020fsei,viola_2023_digital_twins,scarpolini2025nudging}, and therefore we only summarize the main details. 

The fluid dynamics is solved with incompressible (i.e., $\nabla\cdot\mathbf{u}=0$) Navier–Stokes equations:
\begin{equation}\label{eq:NS}
\frac{\partial \mathbf{u}}{\partial t} + (\mathbf{u} \cdot \nabla) \mathbf{u} = -\frac{1}{\rho}\nabla p + \nu \nabla^2 \mathbf{u} + \mathbf{f}\ ,
\end{equation}
where $\mathbf{u}$ and $ p $ are blood velocity and pressure, respectively, $\rho = 1000~\mbox{kg}/\mbox{m}^3$ is blood density, $\nu=4.8\cdot 10^{-6}~\mbox{m}^2/\mbox{s}$ is kinematic viscosity, and $\mathbf{f}$ is the immersed boundary (IB) forcing term that enforces no-slip conditions on cardiac surfaces. The fluid equations are solved with second-order finite differences on a staggered Cartesian grid, and the IB condition is enforced with a moving least squares (MLS) interpolation for accuracy and stability.
{The Eulerian mesh is discretised with $(N_x,N_y,N_z) = (630, 630, 720)$ grid points, corresponding to $(L_x,L_y,L_z) = (0.21, 0.21, 0.24)~\mbox{m}$}. The corresponding spatial and temporal resolutions are $\Delta x = 3.33\cdot 10^{-4}~\mbox{m}$ and $\Delta t = 3\cdot 10^{-6}~\mbox{s}$, respectively. 

The deformable cardiovascular tissues are simulated with a spring-network formulation according to the approach proposed in Ref.~\cite{deTullio2016MLS}. This formulation represents the heart tissues as two-dimensional triangulated meshes with vertices connected by springs. The mass of each mesh triangle is equally distributed among its three nodes.
The elastic potential energy for a spring between two nodes is defined as:
\begin{equation}\label{eq:inplane}
W_e = \frac{1}{2} k_e (l - l_0)^2,
\end{equation}
where $ l $ and $ l_0 $ are the spring lengths in the instantaneous and stress-free configurations, respectively.  
The elastic coefficient $k_e$ is determined from the formula:
\begin{equation}\label{eq:ke}
k_e = \frac{Eh \sum_i A_i}{l^2},
\end{equation}
where $ E $ is Young’s modulus, $ h $ is the membrane thickness, and $ A_i $ is the area of the triangles sharing in common the edge being considered.  {For the aortic valve considered in this work, we have $h=1$~mm and $E=E_p=0.5$~MPa ($E_p$ being the physiological Young modulus), while for the mitral valve the Young modulus is 0.75~MPa~\cite{stradins2004AV_mech_prop,caballero2017bovine_porcine_pericardium_mech_prop}}. The Young elastic modulus of the aortic tissue is 0.4~MPa, for the left ventricle is 93~kPa, and for the left atrium is 50~kPa~\cite{jehl2021_LV_modulus,moireau_2009_filtering}.
Concerning the bending stiffness, the bending energy is given by $W_b = k_b [1 - \cos(\theta - \theta_0)]$, where $ k_b $ is the bending stiffness constant, $\theta$ and $\theta_0$ are the angles between the normal vectors of two neighboring faces sharing an edge in the instantaneous and stress-free configurations, respectively~\cite{deTullio2016MLS}.
The elastic force acting on each node is computed as the derivative of the total elastic energy $W = W_e+W_b$ with respect to the position of the nodes~\cite{deTullio2016MLS}. 

The electrophysiological dynamics of the myocardium, which governs the ionic fluxes across the myocytes and the subsequent muscular contraction, can be modeled by solving a system of partial differential equations describing the spatiotemporal evolution of the transmembrane potential (details can be found in Refs.~\cite{viola2023fsei,viola2020fsei}). 
In the monodomain formulation, the atrial and ventricular myocardium is modeled as a conductive medium coupled with ionic current dynamics and has been shown to accurately model the electrophysiology of healthy cardiac tissue~\cite{clayton2008guide,clayton2011models,viola2020fsei}.

The resulting dynamics for every mesh node having position $ \mathbf{r}^{(i)}$ [see also Fig.~\ref{fig:sketch}, panel (a)] is calculated according to Newton’s law:
\begin{equation}
m_i \frac{d^2 \mathbf{r}^{(i)}(t)}{dt^2} =  \mathbf{f}_i^A+\mathbf{f}_i^I + \mathbf{f}_i^H\ ,
\end{equation}
where $ m_i $ is the nodal mass, $\mathbf{f}_i^A$ is the active tension based on the cellular action potential solved by the electrophysiology solver$, \mathbf{f}_i^I $ and $ \mathbf{f}_i^H $ are elastic and hydrodynamic forces, respectively.
Hydrodynamic loads are obtained via the integration of pressure and viscous stresses from the fluid solver over the immersed surface mesh. 

\subsection{Lagrangian statistics}\label{sec:lagr_turb}
In order to quantitatively investigate the turbulent properties of the blood flow within the left heart, we simulate the dynamics of $\Ntracers = 10^5$ Lagrangian (passive) tracers. The tracers move with the same velocity of the fluid flow so that the velocity of the $j-$th tracer is:
\begin{equation}\label{eq:tracer_vel}
    \mathbf{U}^{(j)}(t)= \mathbf{u}(\mathbf{q}^{(j)}(t),t)\ ,
\end{equation}
where $\mathbf{q}^{(j)}(t)$ represents the spatial coordinates of the $j$-th tracer at time $t$.
From a numerical point of view, since the Navier-Stokes equations are numerically solved on a Cartesian grid, the fluid velocity $\mathbf{u}$ is known only on lattice nodes. Therefore, the tracer velocity $\mathbf{U}$ must be interpolated. This task can be achieved by using a discrete Dirac delta function $\tilde{\Delta}(\mathbf{x})$: $\mathbf{U}^{(j)}(t)= \sum_{\mathbf{x}} \mathbf{u}(\mathbf{x},t)\tilde{\Delta}(\mathbf{q}^{(j)}(t)-\mathbf{x}))\Delta x^3$ (details on the computation of the discrete Dirac delta can be found in Ref.~\cite{peskin2002immersed}). In particular, we employed the three-point discrete Dirac delta~\cite{peskin2002immersed}. The position of the Lagrangian tracer is then integrated with a forward Euler scheme: $\mathbf{q}^{(j)}(t+\Delta t) = \mathbf{q}^{(j)}(t) + \mathbf{U}^{(j)}(t)\Delta t$. Note that the time step used to update the position of the Lagrangian tracer is the same as that used to simulate the Navier-Stokes equations [see Eq.~\eqref{eq:NS}], which has been chosen small enough to avoid spurious contributions to the tracer acceleration coming from the discrete interpolation of the velocity.

Fig.\ref{fig:sketch} illustrates representative examples of tracer motion within the left heart. Panel (a) shows the three-dimensional anatomy of the left heart, including the mitral and aortic valves (shown in a darker color), along with a subset of Lagrangian trajectories. {Note that Fig.\ref{fig:sketch} is meant just as a sketch since the trajectories represent the time evolution of the position, while the geometries of the heart and the valves are taken at a given time (specifically, at the beginning of the systole). All the trajectories shown are synchronous and span four cardiac cycles.} 
Three tracers are highlighted to showcase distinct flow behaviors: the yellow tracer remains confined within the ventricle, following a recirculating path; the purple tracer stays in the ventricle for some time before being ejected into the aorta; the green tracer follows a faster, more direct route from the ventricle to the aortic arch. Insets provide close-up views of segments of the purple and green paths, revealing their intricate trajectories and interactions with local flow structures.
Panels (b-d) show the corresponding time evolution of their velocity components over four cardiac cycles, including $U_x(t)$, $U_y(t)$, $U_z(t)$, and the velocity magnitude $|\mathbf{U}(t)|$. The background shading indicates the chamber where the tracer is located at any given time, helping to link the velocity signals with the tracer’s position in the heart. The letters (e–h) refer to the four panels on the bottom, which display the colormap of the velocity magnitude at four distinct phases of the cardiac cycle. 

\begin{figure}[t!]
    \centering
    \includegraphics[width=1.\linewidth]{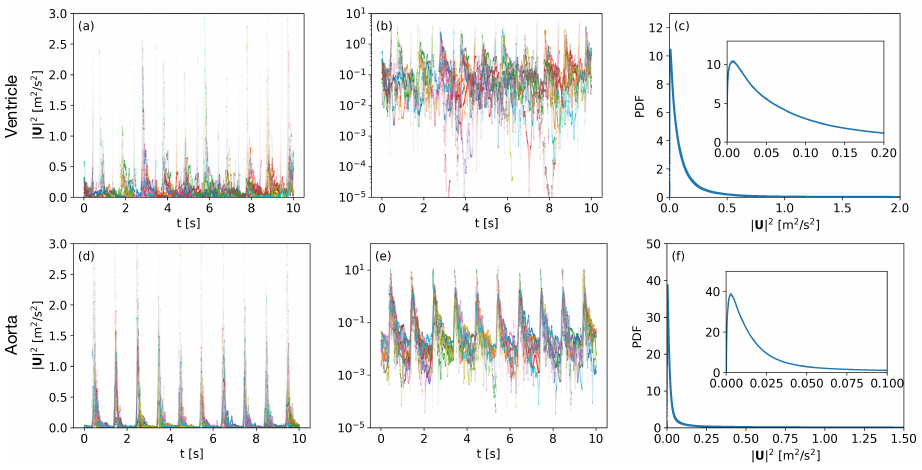}
    \caption{Lagrangian statistics of tracers' squared modulus of velocity $|\textbf{U}|^2$ in the ventricle [panels (a-c)] and in the aorta [panels (d-f)]. Panels (a) and (d) show $|\mathbf{U}(t)|^2$ in linear scale, while panels (b) and (e) report the same quantity in logarithmic scale. Panels (c) and (f) report the corresponding probability distribution function (PDF) with an inset showing a zoomed view at small values.}
    \label{fig:v_a_ventricle}
\end{figure}

Tracer velocity $\textbf{U}^{(j)}(t)$ is then used to compute the main quantities that are typically employed in Lagrangian turbulence to quantitatively assess the turbulent properties of the flow and that are introduced and defined below. 
It is important to note that all the statistical quantities presented in this section are computed separately for each heart chamber. Specifically, tracers are continuously labeled according to the chamber where they reside, allowing for a chamber-specific analysis. This enables us to independently assess the statistics within the ventricle and the aorta [with the further distinction between its ascending and descending parts, see Fig.~\ref{fig:sketch}, panel (a)], as well as to consider their combination, which we will refer to as the ``heart’’.

\subsubsection{2-point correlation function}\label{sec:stat-corr}
A simple quantity to be analyzed is the 2-point correlation function for the Lagrangian kinetic energy  $|\mathbf{U}(t)|^2$:
\begin{equation}\label{eq:conn_corr}
C(\tau) = \frac{\langle |\mathbf{U}(t)|^2 |\mathbf{U}(t - \tau)|^2 \rangle-\langle |\mathbf{U}(t)|^2\rangle \langle |\mathbf{U}(t - \tau)|^2 \rangle}{\langle \left(|\mathbf{U}(t)|^2\right)^2 \rangle-\langle|\mathbf{U}(t)|^2 \rangle^2} \ ,
\end{equation}
where $\tau$ is a time lag; the average $\langle\ \cdot\ \rangle$ is considered over both the set of  Lagrangian tracers $\Ntracers$ and time. {We remark that tracers are labeled depending on the chamber they are in: this means that, when computing the time average for each tracer, we can consider the time interval in which a tracer is in a given chamber.}

The behavior of $C(\tau)$ can be used to distinguish between different flow regimes: in laminar or weakly turbulent flows, energy fluctuations remain correlated over longer timescales, whereas in highly turbulent regimes, correlations decay more rapidly due to increased mixing and energy transfer across scales.
However, it is well known that to fully characterize and distinguish different turbulent realizations, it is necessary to look also at higher orders correlation functions, in order to quantify, scale-by-scale, the presence of deviations from a simple quasi-Gaussian statistics and from pure self-similar behavior. In order to do that, it is customary in Lagrangian turbulence, as for the Eulerian counterpart, to introduce the structure functions and the generalized flatnesses \cite{biferale2008lagrangian,Benzi2023,toschiLagrangianPropertiesParticles2009}.

\subsubsection{Structure functions {and generalized  flatnesses}}\label{sec:stat-sf}
The Lagrangian structure functions of order $p$ over a time lag $\tau$ are defined as~\cite{toschiLagrangianPropertiesParticles2009}:
\begin{equation}
    S_{p,i} (\tau) = \langle[U_i(t)-U_i(t-\tau)]^p\rangle = \langle[\delta_{\tau}U_i(t)]^p\rangle 
\end{equation}
being $i=x,y,z$ the three velocity components. 
Since the system we are studying is neither homogeneous nor isotropic, we consider the following quantity to get rid of the dependence on the spatial directions:
\begin{equation}\label{eq:delta_V}
    \delta_\tau V(t) = \sqrt{\sum_i[\delta_\tau U_i(t)]^2}\ ,
\end{equation}
where the sum runs over the three velocity components. 
The structure function is therefore computed as:
\begin{equation}\label{eq:sp}
S_p(\tau) = \langle(\delta_{\tau}V)^p\rangle \ . 
\end{equation}

\begin{figure}[t!]
    \centering
    \includegraphics[width=1.\linewidth]{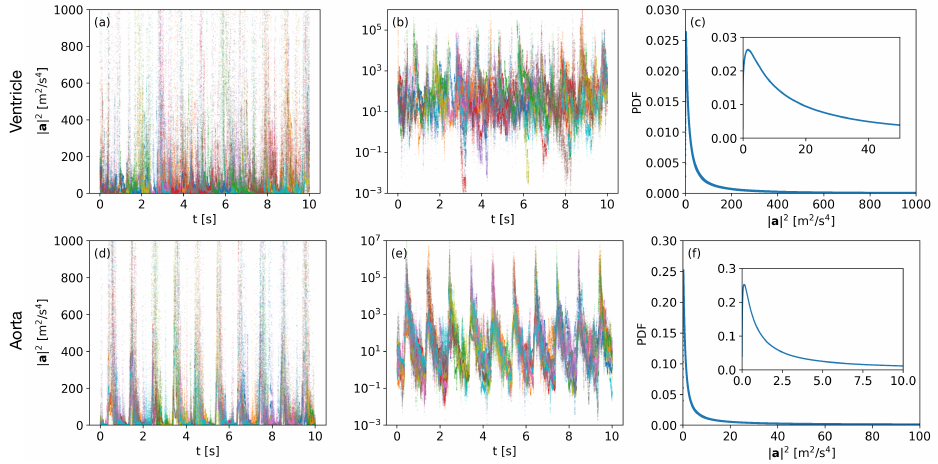}
    \caption{Same as Fig.~\ref{fig:v_a_ventricle}, but for the Lagrangian tracers' squared modulus of acceleration $|\mathbf{a}|^2$.
    }
    \label{fig:v_a_aorta}
\end{figure}

Similarly, one defines the generalized flatnesses of order $p$ for all time lags $\tau$ by the following  dimensionless combination~\cite{Benzi2023}:
\begin{equation}\label{eq:F}
    F_p (\tau) = \frac{S_p(\tau)}{[S_2(\tau)]^{p/2}}.
\end{equation}
Structure functions provide a statistical tool for describing turbulent velocity fluctuations~\cite{toschiLagrangianPropertiesParticles2009}. In the presence of a pure self-similar behavior, the following dimensional relations must hold:
\begin{equation}\label{eq:sp+F}
S_p(\tau) \sim [S_2(\tau)]^{p/2}\quad \Longleftrightarrow \quad  F_p (\tau) \sim \text{const} \ ,\ \forall \tau \ .
\end{equation}
Any deviations from the above behaviors must be connected to the presence of non-trivial scale-by-scale fluctuations that cannot be captured by the 2-point correlation function only~\cite{benziExtendedSelfsimilarityTurbulent1993}. Deviation from self-similarity also implies the impossibility of collapsing the probability density function (PDF) of the standardized velocity increments $\delta_\tau V(t)/
\langle    [\delta_\tau V(t)]^2 \rangle^{1/2}$. 
We notice that the $p$-th  moments of the latter quantity coincide with the generalized flatnesses [Eq.~\eqref{eq:F}].

{In order to quantitatively assess any deviation from pure self-similar scaling using quantities that stay  $\mathcal{O}(1)$ for all time increments, it is customary to redefine the structure functions in terms of their ``local exponents''~\cite{Benzi2023, biferale2008lagrangian}:
\begin{equation}\label{eq:xi}
\xi_p(\tau) = \frac{d\log S_p(\tau)}{d\log \tau} \ .
\end{equation}
In ideal homogeneous and isotropic turbulence, it is known that the local exponents become $\tau$-independent in the inertial range,
for time lags much smaller than the large-scale characteristic correlation time and much larger than the viscous dissipative time~\cite{benziExtendedSelfsimilarityTurbulent1993}. However, in the inertial range of fully developed turbulence, they are observed to assume anomalous values, different from the $\xi_p = p/2$ dimensional prediction given  by the Lagrangian equivalent of the K41 theory~\cite{benziExtendedSelfsimilarityTurbulent1993,biferale2008lagrangian}. On the other hand, for all flows in Nature, we expect to observe the differentiable limit, i.e., $\xi_p(\tau) \rightarrow p$, for time increments $\tau\to 0$.}

{The importance of the local exponents [Eq.~\eqref{eq:xi}] goes much beyond their impact on homogeneous and isotropic turbulence. They define a dimensionless metric for the deviation from self-similarity and a quantitative tool to distinguish different turbulent flows, scale-by-scale, without using log-log fits or log-log plots~\cite{benziExtendedSelfsimilarityTurbulent1993}. In particular, it is easy to recognize that we can exactly rewrite the expression for Eq.~\eqref{eq:F} as:
\begin{equation}\label{eq:F2}
    F_p (\tau) = S_2(\tau)^{\chi_p(\tau)} \ ,
\end{equation}
with  
\begin{equation}\label{eq:xipxi2}
\chi_p(\tau) = \frac{\xi_p(\tau)}{  \xi_2(\tau)}-\frac{p}{2} \ ,
\end{equation}
that gives a direct measurement of the anomalous scale-by-scale dependency of the standardized velocity increment statistics. In other words, in a purely self-similar field,  the dimensional relation given by Eq.~\eqref{eq:sp+F} would hold, and $\xi_p(\tau) = p/2$, leading to $F_p(\tau) = \text{const}, \forall \tau$.  It is important to remark that the metrics given in Eqs.~\eqref{eq:F2} and~\eqref{eq:xipxi2} are well defined and meaningful even in the presence of strongly ``non-ideal'' turbulent statistics, where power laws and pure fractal or multifractal behaviors do not hold because of anisotropic and non-homogeneous boundary conditions. From a practical perspective, analyzing $\chi_4(\tau)$ in different regions of the cardiovascular system will also help to distinguish turbulence characteristics between heart chambers.}

\begin{figure}[t!]
    \centering
    \includegraphics[width=1.\linewidth]{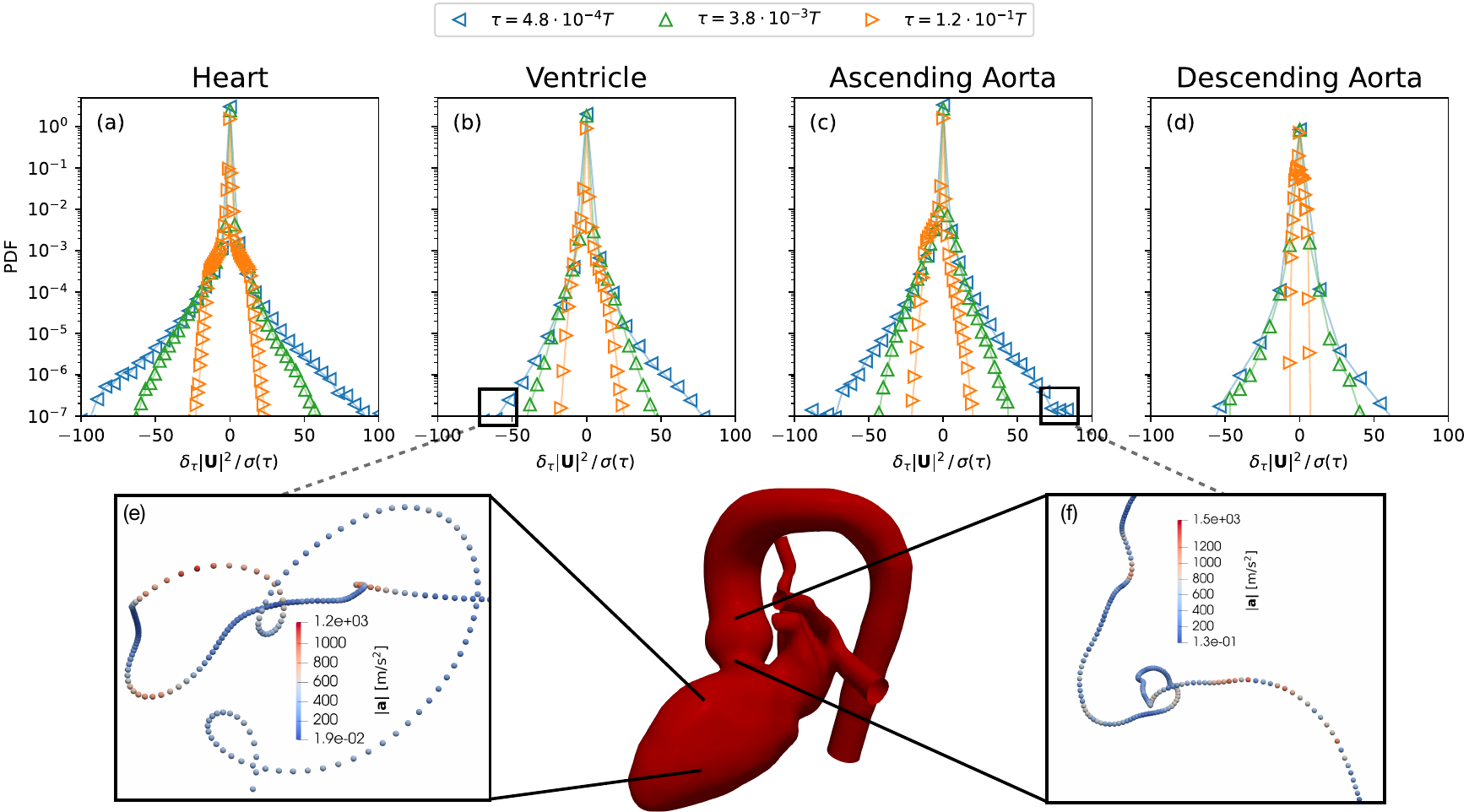}
    \caption{Top panels (a-d): probability distribution functions (PDFs) of the normalized kinetic energy increments, $\delta_\tau |\mathbf{U}|^2 / \sigma(\tau)$, for three time delays: $\tau = 4.8 \cdot 10^{-4} T$ (blue), $3.8 \cdot 10^{-3} T$ (green), and $1.2 \cdot 10^{-1} T$ (orange). Results are shown separately for tracers in the heart (a), ventricle (b), ascending aorta (c), and descending aorta (d). {Bottom panels (e-f): representative tracer trajectories associated with extreme events ($\delta_\tau |\mathbf{U}|^2 / \sigma(\tau) > 50$ at $\tau = 4.8 \cdot 10^{-4} T$) are shown in the ventricle [panel (e)] and in the ascending aorta [panel (f)]. The central panel displays the anatomical model of the heart and indicates the locations of the zoomed-in views.}
    Simulations correspond to HR = 60 and valve Young modulus $E = E_p$.}
    \label{fig:pdf1-chamb}
\end{figure}
\begin{figure}[t!]
    \centering
    \includegraphics[width=1.\linewidth]{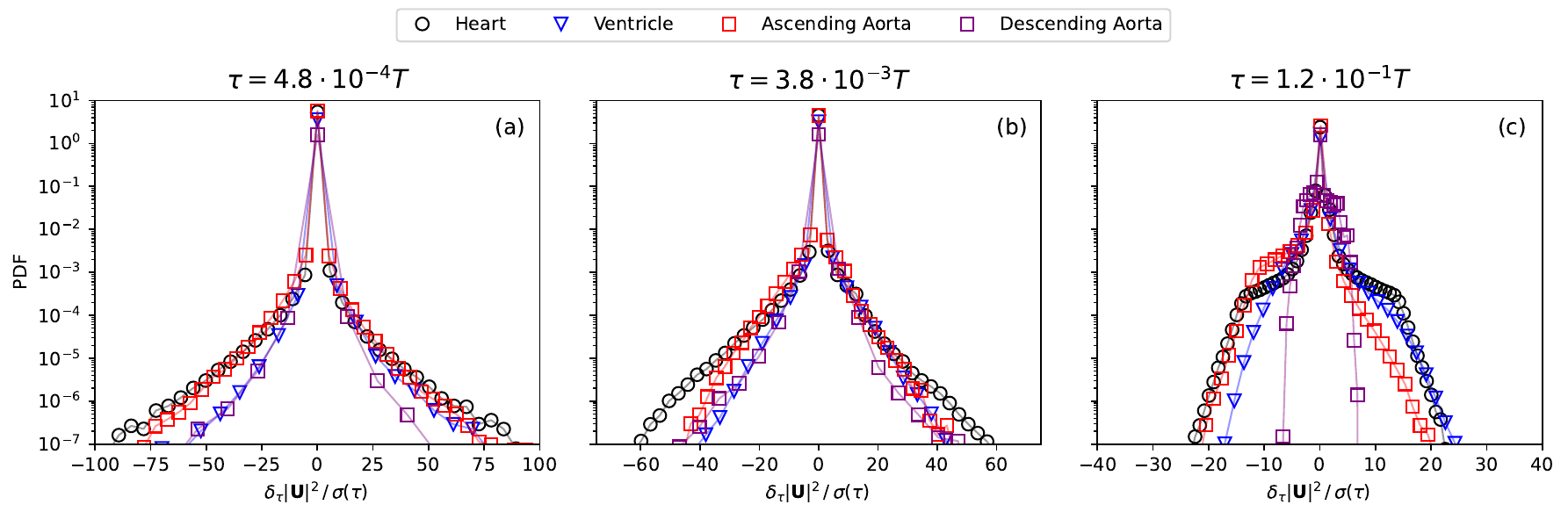}
    \caption{Comparison of the probability distribution functions (PDFs) of the normalized kinetic energy increments, $\delta_\tau |\mathbf{U}|^2 / \sigma(\tau)$, across different chambers [heart (black circles), ventricle (blue triangles), ascending aorta (red squares), and descending aorta (purple squares)] for fixed time delays: $\tau = 4.8 \cdot 10^{-4} T$ (a), $3.8 \cdot 10^{-3} T$ (b), and $1.2 \cdot 10^{-1} T$ (c). Simulations correspond to HR = 60 and valve Young modulus $E = E_p$.}
    \label{fig:pdf2-chamb}
\end{figure}

\section{Results}\label{sec:results}

In this section, we discuss the main statistical quantities measured from the numerical simulations. First, we focus on quantitative assessments of turbulent fluctuations across different chambers (see Sec.~\ref{sec:results-chamb}), and then we briefly discuss the sensitivity of such measures at varying the heart rate HR and the stiffness of the leaflets of the aortic valve (see Sec.~\ref{sec:results-HR} and Sec.~\ref{sec:results-ke}, respectively).

For all the above-mentioned cases, we simulate the evolution of the Lagrangian tracers for ten cardiac cycles. For each of them, we probe the velocity $\vec{U}(t)$ every $\delta  t=10\Delta t=3\cdot 10^{-5}\mbox{ s}$. {The choice of $\delta t$ is such that its value is small enough to ensure a good resolution for the smallest temporal scales.}
To perform a detailed statistical analysis for each chamber, a ray-tracing algorithm is used to distinguish between different regions of the heart (ventricle, ascending aorta and descending aorta). Tracers are then labeled according to their location, enabling a more precise evaluation of turbulence characteristics within each chamber.
All the tracers are randomly initialized in the left atrium with the initial velocity given by Eq.~\eqref{eq:tracer_vel}. During the whole simulation, the number of tracers is fixed to $\Ntracers$: this means that as soon as a tracer exits from the descending aorta, a new tracer is spawned at a random location within the left atrium. 

We first analyze simulations at fixed heart rate (HR=60~bpm) and physiological aortic valve stiffness ($E=E_p$, see Sec.~\ref{sec:numerical} and Eqs.~\eqref{eq:inplane} and~\eqref{eq:ke}). Then, we compare two simulations at different heart rates within the physiological range (HR~=~40~bpm and 80~bpm) and $E=E_p$. Finally, we investigate the effect of valve stenosis (mimicked by an enhanced elastic stiffness of its leaflets) for the baseline HR = 60 bpm.

\begin{figure}[t!]
    \centering    \includegraphics[width=.8\linewidth]{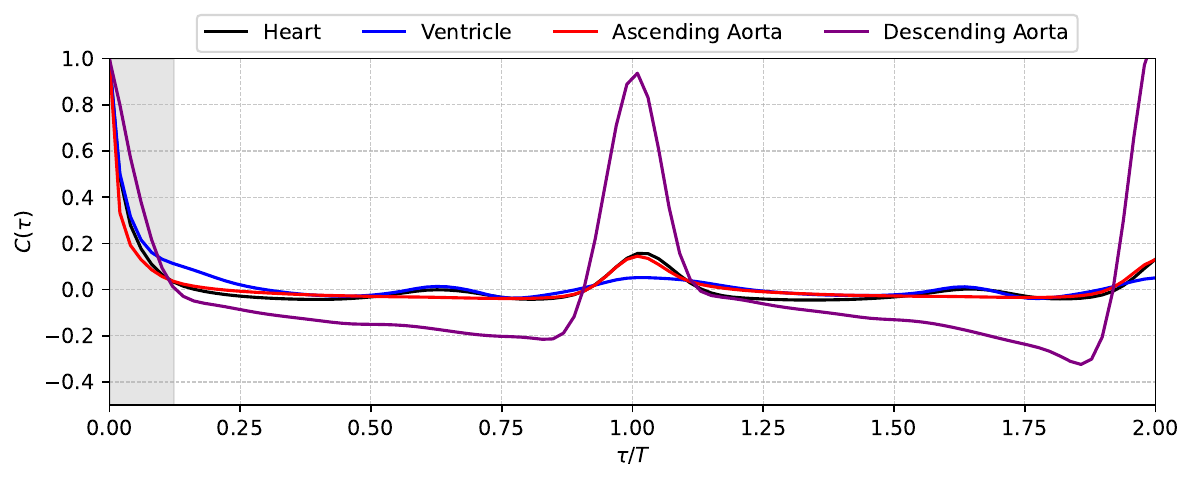}
    \caption{2-point correlation function $C(\tau)$ [see Eq.~\eqref{eq:conn_corr}] of the kinetic energy $|\mathbf{U}|^2$ for tracers in the heart (black), ventricle (blue), ascending aorta (red) and descending aorta (purple). Time lag $\tau$ is normalized by the cardiac period $T$. Results are shown for HR = 60~bpm and valve Young module $E = E_p$.}
    \label{fig:corr-chamb}
\end{figure}

\subsection{Statistical analysis across chambers}\label{sec:results-chamb}
Fig.~\ref{fig:v_a_ventricle} and Fig.~\ref{fig:v_a_aorta} present a comparison of the Lagrangian statistics of squared modulus of velocity ($|\textbf{U}|^2$, Fig.~\ref{fig:v_a_ventricle}) and acceleration ($|\textbf{a}|^2$, Fig.~\ref{fig:v_a_aorta}) for tracers in the ventricle and aorta. In both figures, {twenty individual Lagrangian trajectories are shown, each reported in a different color to highlight the variability across tracers}. The left and center panels show their time evolution over a 10-second interval (corresponding to ten cardiac cycles, as the heart rate is HR = 60 bpm), while the right panels display the corresponding probability distribution functions (PDFs).
From Fig.~\ref{fig:v_a_ventricle}, we can see that the time evolution of $|\textbf{U}|^2$ in the ventricle [panels (a) and (b)] displays intermittent behavior: most of the time, the values of $|\mathbf{U}|^2$ are small, interrupted by sharp spikes associated with systolic ejection and mitral inflow. In contrast, the values of $|\textbf{U}|^2$ in the aorta [panels (d) and (e)] reveal a more regular, periodic pattern, with distinct peaks corresponding to systolic ejection followed by quieter phases. These peaks are smoother and more synchronized across tracers, reflecting the more uniform and directional flow in the aorta. Moreover, the ventricle is also subjected to the mitral inflow, which reduces the duration of the resting phase.
The time evolution of $|\textbf{a}|^2$ (Fig.~\ref{fig:v_a_aorta}) shows some similar features, with strong peaks capturing abrupt velocity changes: both in the ventricle and aorta, the values of the acceleration can reach even 100 times the acceleration of gravity, as typical in turbulent flows~\cite{la2001fluid,toschiLagrangianPropertiesParticles2009}.
The PDFs of both $|\mathbf{U}|^2$ [Fig.~\ref{fig:v_a_ventricle}, panels (c) and (f)] and $|\mathbf{a}|^2$ [Fig.~\ref{fig:v_a_aorta}, panels (c) and (f]) are highly skewed and exhibit fat tails, clear markers of strong Lagrangian intermittency due to the complex, time-varying flow structures. This semi-quantitative comparison of a few tracers suggests that while both regions (ventricle and aorta) experience intermittent Lagrangian dynamics, the nature and intensity of intermittency differ. 

\begin{figure}[t!]
    \centering
    \includegraphics[width=1.\linewidth]{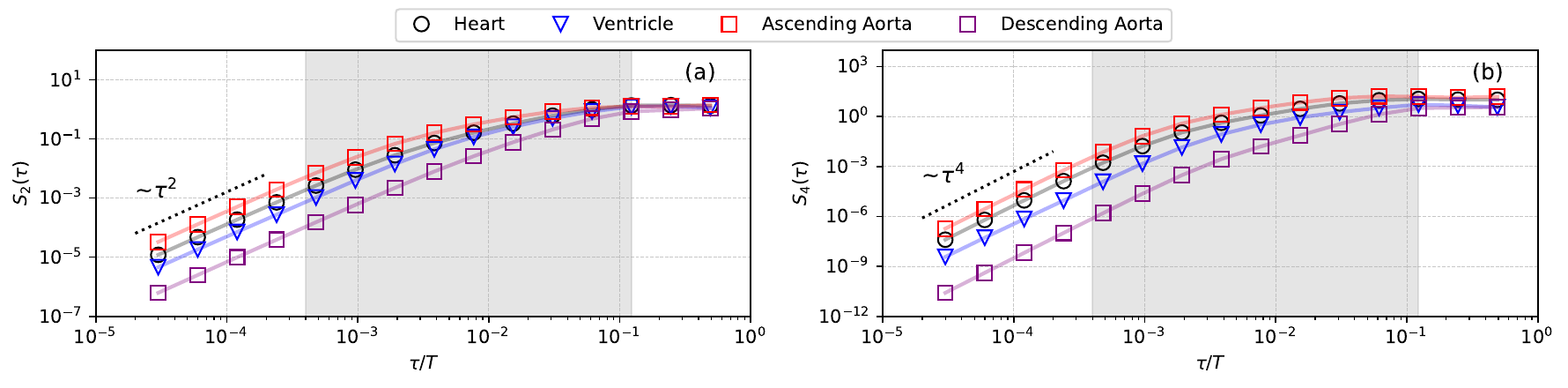}
    \caption{Second- and fourth-order structure functions [$S_2(\tau)$, panel (a), and $S_4(\tau)$, panel (b), respectively, see  Eq.~\eqref{eq:sp}] at different scales for the heart (black circles), ventricle (blue triangles), ascending aorta (red squares) and descending aorta (purple squares). Data for HR = 60~bpm and $E = E_p$.}
    \label{fig:sf-chamb}
\end{figure}
\begin{figure}[t!]
    \centering
    \includegraphics[width=.9\linewidth]{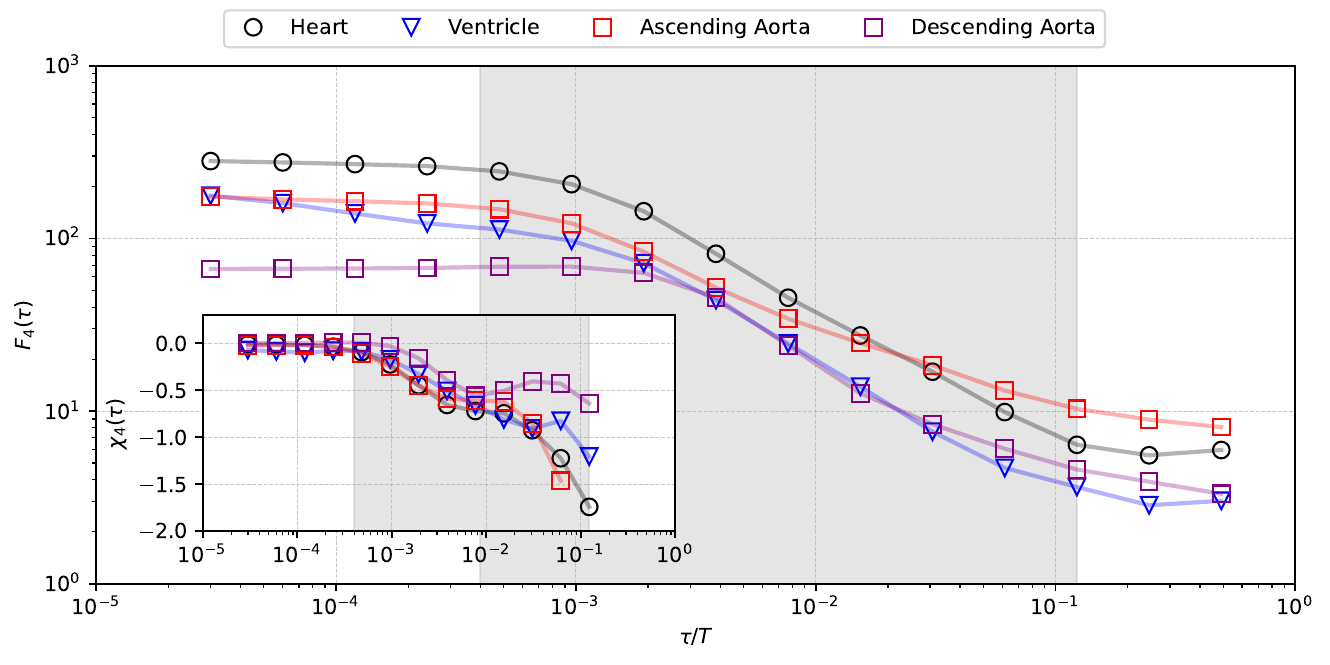}
    \caption{{Fourth-order flatness $F_4(\tau)$ [see Eq.~\eqref{eq:F}]} at different scales for the heart (black circles), ventricle (blue triangles), ascending aorta (red squares) and descending aorta (purple squares). The inset shows $\chi_4(\tau)$ [see Eq.~\eqref{eq:xipxi2}].
    Data for HR = 60~bpm and $E = E_p$.
    }
    \label{fig:F4-chamb}
\end{figure}

{Before entering into the details of the statistical quantities presented in Sec.~\ref{sec:lagr_turb}, we start analyzing the PDFs of kinetic energy increments, $\delta_\tau|\mathbf{U}|^2$, normalized by their variance $\sigma(\tau)$. Results are reported in 
Fig.~\ref{fig:pdf1-chamb}. The statistics have been collected across the heart [panel (a)], the ventricle [panel (b)], the ascending aorta [panel (c)], and the descending aorta [panel (d)]. {We remark that, in this section, when we refer to ``heart'' we mean ventricle and aorta together.} 
The PDFs are computed for three different time lags $\tau$, that have been chosen according to the behavior of the structure functions (more details are provided below, see Fig.~\ref{fig:sf-chamb}): the smallest and largest values of $\tau$ (i.e., $\tau=4.8\cdot 10^{-4}T$ (blue) and $\tau=1.2\cdot 10^{-1}T$ (orange)) identify the range where the most relevant physical mechanisms of the turbulent cascade are active, while $\tau=3.8\cdot 10^{-3}T$ (green) represents an intermediate value. $T$ represents the cardiac period.}
The PDFs reveal clear evidence of scale-dependent intermittency, with deviations from Gaussian statistics becoming more pronounced as $\tau$  decreases. 
The distributions exhibit fat tails at the smallest $\tau$, indicating that extreme velocity fluctuations are significantly more probable than in a Gaussian process. This is a key signature of strong intermittency, where turbulent energy transfer mechanisms produce rare but intense velocity fluctuations.
{Panels (e) and (f) report some representative trajectories that had some extreme events that contributed to the tails in the ventricle and in the ascending aorta, respectively. It is also worth noting that the tails in the descending aorta [panel (d)] are less pronounced than those in the ascending aorta [panel (c)].}
To further explore these differences across chambers, Fig.~\ref{fig:pdf2-chamb} presents the same PDFs but reorganized to compare different chambers at fixed $\tau$ values.
Comparing the ventricle (blue triangles) and the ascending aorta (red squares), we observe a distinct difference in intermittency levels. The ascending aorta generally exhibits stronger fat tails than the ventricle, suggesting a higher degree of intermittency and non-Gaussian statistics. This behavior is likely due to the intense systolic ejection, which produces high-velocity fluctuations. Panel (c) shows a distinct bump in the negative tail of the distribution for the ascending aorta, indicating an excess of large negative fluctuations in $|\mathbf{U}|^2$ at this time delay ($\tau=1.2\cdot 10^{-1}T$). This behavior likely reflects strong deceleration events that occur after systolic ejection, when the high-speed outflow is abruptly slowed down [as can also be seen in Fig.~\ref{fig:v_a_ventricle}, panels (d-e)]. These decelerations are both coherent and recurrent, leading to a noticeable, non-Gaussian enhancement in the left tail of the distribution.
For the heart (black), bumps are visible in both tails of the distribution. 
The bump on the right may correspond to positive increments caused by tracer acceleration during ventricular ejection: in the ventricle, such a bump is not present because as soon as a tracer exits the ventricle, it is not considered anymore in the statistics of the ventricle itself. The bump on the left likely reflects the deceleration processes that take place in the aorta. {As already noticed, the descending aorta (purple squares) shows a lesser degree of intermittency with respect to the ascending aorta.} The qualitative trends shown by the standardized PDF analyzed in this section are a first important indication that different non-trivial multi-scale properties are enjoyed in different heart chambers and under different heart conditions. In the following section, we quantify these trends by using high-order structure functions and generalized flatnesses.
We note that, in this section, we chose to analyze the PDFs of the Lagrangian kinetic energy $|\mathbf{U}(t)|^2$ rather than the velocity increments $\delta_\tau V(t)$ [which will be used in the next section for computing structure functions, see Eq.~\eqref{eq:delta_V}] because the latter is a positive-definite quantity, which yields a chi-squared-like distribution rather than a Gaussian-like one.

\subsubsection{Scale-by-scale analysis}
Fig.~\ref{fig:corr-chamb} shows the 2-point correlation function $C(\tau)$ of $|\textbf{U}|^2$ [see Eq.~\eqref{eq:conn_corr}] in the heart (black line), ventricle (blue line), ascending aorta (red line) and descending aorta (purple line) as a function of normalized time delay $\tau/T$.
The gray-shaded region has been selected based on the behavior of the structure functions ($S_2(\tau)$ and $S_4(\tau)$) and $\chi_4(\tau)$ given in Eq.~\eqref{eq:xipxi2} (see later Figs.~\ref{fig:sf-chamb} and~\ref{fig:F4-chamb}).

The correlation function in the {ascending} aorta decays rapidly to zero, indicating that velocity fluctuations in this region are uncorrelated beyond a characteristic timescale comparable to the systole duration, which in our simulations is $\sim 0.2$~s. This behavior can be due to the strong advection of flow in the aorta, where turbulence is primarily driven by pulsatile forces and lacks long-term memory.
{In contrast, the correlation function in the descending aorta exhibits a slower decay and a prolonged range of anti-correlation; hence, for a time delay comparable to the cardiac period ($\tau/T \sim 1$), the correlation remains stronger than in the other chambers. This behavior may result from both the lower velocity magnitudes observed in the descending aorta compared to the other regions and the reduced turbulence intensity, as already indicated by the PDFs (see Figs.~\ref{fig:pdf1-chamb} and~\ref{fig:pdf2-chamb}). The presence of both a pronounced correlation around $\tau \sim T$ and a persistent anti-correlation over a wide range of $\tau/T$ suggests that the flow structures in the descending aorta are more coherent and less disordered than those in the other chambers.}
The correlation function in the ventricle does not decay immediately to zero, suggesting the presence of coherent structures that persist over time. This long-lived correlation reflects the cyclic nature of ventricular contraction and relaxation, which continuously influences the velocity field. While the correlation function for the ascending aorta flattens after reaching zero and remains close to zero until a time-lag $\tau$ comparable to the period of the heartbeat $T$, the ventricle exhibits a secondary peak at $\tau/T \sim 0.6$. This feature may be attributed to the coexistence of systolic ejection and mitral inflow, which introduce two distinct characteristic timescales in the ventricular flow dynamics.
The differences in $C(\tau)$ between the ascending aorta and the ventricle highlight the distinct flow regimes in these regions: while the ascending aorta exhibits rapid decorrelation due to strong advection and turbulence dissipation, the ventricle retains memory of its flow through recirculations and cyclic contraction dynamics.

\begin{figure}[t!]
    \centering
    \includegraphics[width=1.\linewidth]{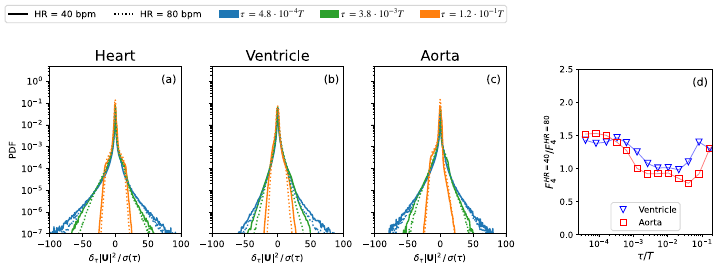}
    \caption{Effect of heart rate on Lagrangian statistics. Panels (a-c) show probability distribution functions (PDFs) of the normalized kinetic energy increments, $\delta_\tau |\mathbf{U}|^2 / \sigma(\tau)$, for three time delays: $\tau = 4.8 \cdot 10^{-4} T$ (blue), $3.8 \cdot 10^{-3} T$ (green), and $1.2 \cdot 10^{-1} T$ (orange). Results are shown separately for tracers in the heart (a), ventricle (b), and aorta (c).
    Data are reported for HR = 40~bpm (solid lines) and HR = 80~bpm (dashed lines).
    Panel (d) shows the ratio between the fourth-order flatness at HR = 40 and HR = 80, $F_4^{{\mbox{\tiny HR=40}}}(\tau)/F_4^{{\mbox{\tiny HR=80}}}(\tau)$, as a function of the normalized time delay $\tau/T$, for tracers in the ventricle (blue triangles) and aorta (red squares). Valve stiffness is fixed at $E = E_p$.}
    \label{fig:sf-HR}
\end{figure}

{As already discussed in Sec.~\ref{sec:lagr_turb}, the correlation function alone is insufficient to fully characterize the turbulent features of the flow {due to the presence of strong multi-scale properties as qualitatively shown by the PDF’s behavior in the previous figures}. To gain deeper insight, we extend our Lagrangian statistical analysis to include the structure functions $S_2(\tau)$ and $S_4(\tau)$ [see Eq.~\eqref{eq:sp}].  
Fig.~\ref{fig:sf-chamb} shows the statistical behavior of the Lagrangian velocity increment magnitude $\delta_\tau V$ [defined in Eq.~\eqref{eq:delta_V}] across multiple scales, using  $S_p(\tau)$ for $p = 2$ [panel (a)] and $p = 4$ [panel (b)].}
{As for the correlation function,} the statistic is conditioned across different chambers [ventricle (blue triangles), ascending aorta (red squares) and descending aorta (purple squares)] as well as in the heart (black circles).
The gray-shaded region has been selected as the region where the most relevant turbulent aspects are shown.
Indeed, on the one hand, for $\tau/T \lesssim 4\cdot 10^{-4}$, the signal is smooth and differentiable, and the scaling $S_p(\tau) \sim \tau^p$ for small values of $\tau$ is observed (see Sec.~\ref{sec:stat-sf}); on the other hand, for $\tau/T \gtrsim 1.2\cdot 10^{-1}$, the structure functions become flat, which means that the signals are not correlated. Note that the left extreme of this region also gives an estimate of the maximum $\Delta t$ that can be used in the numerical integration of the Navier-Stokes equations [see Eq.~\eqref{eq:NS}], that is, $\Delta t_{\mbox{\tiny{MAX}}}\sim 0.4$~ ms.
Concerning the comparison across chambers, the structure functions computed in the ascending aorta show larger values than those in the ventricle and in the descending aorta, suggesting stronger velocity fluctuations. 

Fig.~\ref{fig:F4-chamb} shows the fourth-order flatness $F_4(\tau)$ as a function of the normalized time lag $\tau/T$, which provides a measure of intermittency and breaking of self-similarity in the same time range highlighted with a gray region before.  
As further evidence that the descending aorta is less turbulent than the ascending part, we observe that for small values of $\tau/T$, the flatness $F_4(\tau)$ remains nearly constant up to $\tau/T \sim 10^{-3}$, whereas in the ascending aorta, $F_4(\tau)$ already exhibits a scale dependence around $\tau/T \sim 10^{-4}$.

As discussed in Sec.~\ref{sec:stat-sf}, the degree of intermittency can be further characterized across different time scales using the ratio $\xi_4(\tau) / \xi_2(\tau)$. In Fig.~\ref{fig:F4-chamb}, inset, we observe that the quantity $\chi_4(\tau)=\xi_4(\tau)/ \xi_2(\tau)-2$ varies between chambers.  
For small values of $\tau$, $\chi_4(\tau)$ remains relatively close to 0 due to the smooth differentiable limit achieved for $\tau \to 0$, where $\xi_p \to p$.  
As $\tau$ increases, $\chi_4(\tau)$ decreases consistently, indicating the emergence of stronger intermittency effects. This trend reflects increasing deviations from a self-similar scaling as rare and intense velocity fluctuations become more dominant. 
{The ventricle, the ascending aorta, and the statistics in the heart exhibit similar trends. In contrast, the descending aorta maintains values of $\chi_4(\tau)$ closer to zero, suggesting a reduced degree of intermittency and a flow structure that is more persistent and less turbulent.}

\subsection{Statistical analysis at varying heart rate}\label{sec:results-HR}
We are now interested in exploring how the heart rate within the healthy range influences Lagrangian statistics. To this end, we compare two simulations performed at HR = 40~bpm and HR = 80~bpm. {Since the contribution of the descending aorta was minor, in this section, we consider the aorta as a whole, i.e., the combination of its ascending and descending parts.} 

{Fig.~\ref{fig:sf-HR}, panels (a–c), shows the PDFs of the normalized kinetic energy increments, $\delta_\tau |\mathbf{U}|^2 / \sigma(\tau)$, computed for three different time delays. The panels correspond to the entire heart, the ventricle, and the aorta, respectively. Results for a heart rate of 40~bpm are shown as solid lines, while those for 80~bpm are indicated with dotted lines. 
We observe, in general, that the case HR = 40 bpm shows slightly fatter tails, indicating a higher degree of intermittency -- this is especially true for the smaller values of $\tau$. This is also confirmed by the study of the flatness $F_4(\tau)$.}
Indeed, in Fig.~\ref{fig:sf-HR}, panel (d), we report the ratio between the fourth-order flatness computed at the two heart rates, $F_4^{{\mbox{\tiny HR=40}}}(\tau)/F_4^{{\mbox{\tiny HR=80}}}(\tau)$, plotted as a function of the normalized time delay $\tau/T$. Across almost all scales and in both chambers, the ratio is greater than one, indicating that the lower heart rate is associated with higher levels of intermittency. The slightly increased intermittency at lower HR suggests that the longer cardiac cycle provides more time for coherent flow structures to develop and persist, leading to stronger velocity fluctuations and more extreme events. 
{It is important to notice that the main difference between the two cases is the duration of the diastasis, i.e., the resting period between the rapid diastolic filling and the atrial systole~\cite{meschini2020heart}.}
Overall, we can conclude that the effect of the heart rate (when varying in the normal range) on the cardiac turbulent features is not marked.

\begin{figure}[t!]
    \centering
    \includegraphics[width=0.8\linewidth]{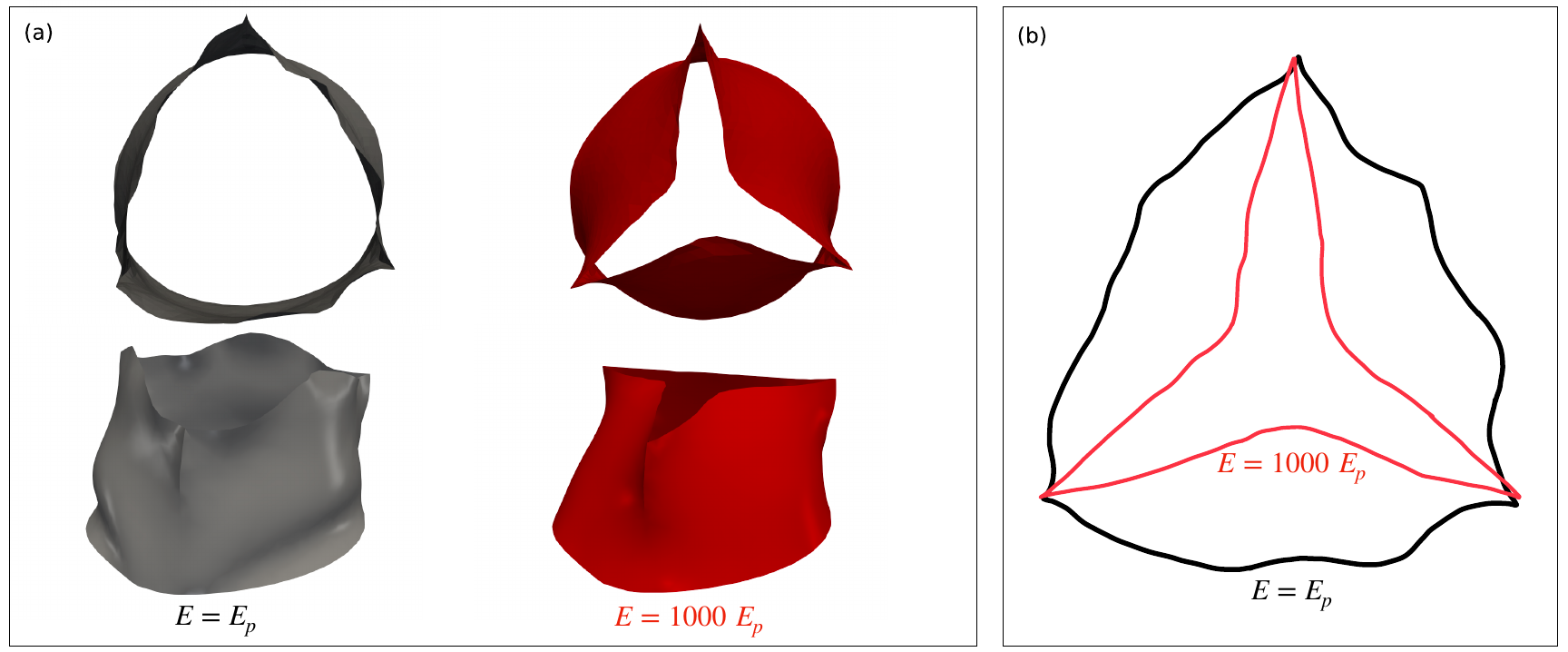}
    \caption{Effect of aortic valve stiffness on leaflet deformation during peak systole. Panel (a): instantaneous configuration of the aortic valve leaflets for two different levels of elasticity: physiological compliance ($E = E_p$, left) and a stiffened valve ($E = 1000\ E_p$, right). The top views illustrate the valve opening area, while the bottom views emphasize the three-dimensional deformation of the leaflets. 
    Panel (b): cross-sectional contour of the valve orifice at peak systole for both cases. The black curve corresponds to the healthy valve, and the red curve to the stenotic one.}
    \label{fig:sketch-ke}
\end{figure}
\begin{figure}[th!]
    \centering
    \includegraphics[width=1.\linewidth]{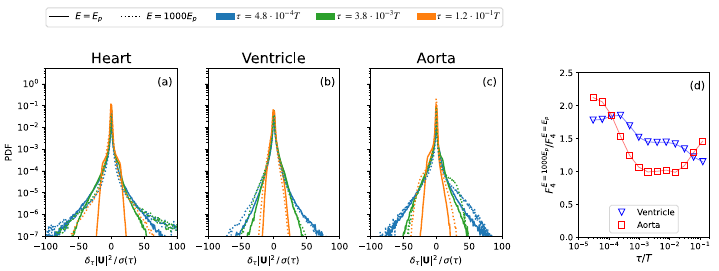}
    \caption{Effect of aortic valve stiffness on Lagrangian statistics. Panels (a-c) show probability distribution functions (PDFs) of the normalized kinetic energy increments, $\delta_\tau |\mathbf{U}|^2 / \sigma(\tau)$, for three time delays: $\tau = 4.8 \cdot 10^{-4} T$ (blue), $3.8 \cdot 10^{-3} T$ (green), and $1.2 \cdot 10^{-1} T$ (orange). Results are shown separately for tracers in the heart (a), ventricle (b), and aorta (c). 
    Data are reported for $E=E_p$ (solid lines) and $E=1000\ E_p$ (dotted lines).
    Panel (d) shows the ratio between the fourth-order flatness at $E = 1000\ E_p$ (stenotic valve) and $E=E_p$ (healthy valve), $F_4^{E=1000E_p}(\tau)/F_4^{E=E_p}(\tau)$, as a function of the normalized time delay $\tau/T$, for tracers in the ventricle (blue) and aorta (red). Heart rate is fixed at HR = 60~bpm.}
    \label{fig:sf-ke}
\end{figure}

\subsection{Statistical analysis {of a stenotic valve}}\label{sec:results-ke}
{In this section, we analyze the Lagrangian dynamics and turbulent features of the cardiac hemodynamics in the presence of a stenotic aortic valve, which is modeled by increasing the Young modulus $E$ of the valve leaflets (see Sec.\ref{sec:numerical}), making them significantly stiffer and thereby reducing the valve opening during systole (see Fig.~\ref{fig:sketch-ke}). We compare two simulations with the same heart rate (HR = 60 bpm) and anatomical configuration: the reference case, with physiological leaflet stiffness ($E = E_p$), and the stenotic case, where the stiffness is increased to $E = 1000\ E_p$, mimicking a pathological condition in which the valve leaflets are partially calcified~\cite{smid2024novel}. As for the previous case, we consider the aorta as a whole, without distinguishing between ascending and descending parts.} 

{In Fig.~\ref{fig:sf-ke}, panels (a–c), we report the PDFs of the normalized  {kinetic energy} increments, $\delta_\tau |\mathbf{U}|^2 / \sigma(\tau)$, computed for three different time delays. The panels correspond to the entire heart, the ventricle, and the aorta, respectively. Results for a physiological Young modulus $E=E_p$ are shown as solid lines, while those for the stenotic aortic valve ($E=1000E_p$) are reported with dotted lines. 
We can, in general, observe slightly fatter tails for $E=1000E_p$ for small values of $\tau/T$, in particular in the aorta [panel (c)] for positive values of $\delta_\tau |\mathbf{U}|^2$. This effect is related to the increased velocity during the systolic jet due to the decreased valve opening (see Fig.~\ref{fig:sketch-ke}), showing that the PDFs can capture such flow properties. 
To get more insights on the turbulent features, we analyze the fourth-order flatness for both cases:} Fig.~\ref{fig:sf-ke}, panel (d), shows the ratio of the fourth-order flatness computed for the stenotic valve to that of the healthy valve, plotted as a function of $\tau/T$. Overall, the data reveal a clear and significant increase in flatness with the stenotic valve, both at small and large time scales. This indicates a stronger degree of intermittency, consistent with sharper velocity gradients and more intense fluctuations, likely a consequence of the reduced valve opening area. 
This leads to a higher effective Reynolds number during systole, which contributes to enhanced turbulence intensity and stronger intermittency in the flow field.
In the aorta, this effect is most evident at short and long-time delays, while at intermediate scales, the increase in flatness is more modest, with an intermediate region where the ratio is $\sim 1$.
{Additionally, although the most pronounced changes occur in the aorta at small time lags, the ventricle also shows a noticeable increase in flatness, indicating that the increased outflow velocity and the altered jet dynamics affect the ventricular flow structures as well. Moreover, due to the sharper velocity gradients associated with the stiffer valve, the time scale at which the flow becomes differentiable is shifted toward smaller $\tau$, as can be seen from the red curve for small values of $\tau/T$ in Fig.~\ref{fig:sf-ke}, panel (d). This implies that a smaller time step ($\Delta t$) is required in the simulations of the pathological case to ensure numerical precision.}

\section{\label{sec:conclusions}Conclusions}
In this study, we present a comprehensive Lagrangian analysis of turbulent blood flow in the human left heart using high-fidelity numerical simulations based on a patient-specific anatomical model. 
Employing a fully coupled fluid–structure–electrophysiology interaction (FSEI) framework, we tracked passive Lagrangian tracers to characterize the statistical properties of velocity fluctuations.

In our analysis, we compare flow statistics in the ventricle and aorta under physiological conditions (heart rate HR = 60 bpm and a physiological aortic valve). 
Both the ventricle and aorta display strong intermittency with non-Gaussian velocity statistics. The aorta is characterized by more organized, advective dynamics, with intermittent features primarily driven by the sharp transitions associated with systolic ejection and subsequent deceleration. In general, the aortic (and in particular its ascending part) hemodynamics manifests stronger turbulence than the ventricle.
Using a combination of 2-point correlation functions, Lagrangian structure functions, flatness factors, and local scaling exponents, we quantify how turbulent energy transport varies across scales. 
Temporal correlation functions further highlight chamber-specific memory effects, including a secondary peak in the ventricle associated with the cyclical interplay between ejection and inflow phases. However, we show that the correlation function is not enough to fully characterize the turbulent features. On the other hand, the flatness, together with the ratio of local slopes, $\xi_4 / \xi_2$, reveals scale-dependent deviations from self-similar behavior and allows a quantitative study of the turbulence. 

Next, we investigate the impact of heart rate on Lagrangian turbulence characteristics. At a lower heart rate (HR = 40 bpm), we observe a slightly enhanced intermittency, both in the ventricle and aorta, as evidenced by increased flatness values. This counterintuitive enhancement can be attributed to the longer diastatic phase, which allows for the development and persistence of coherent flow structures. 

Nevertheless, varying the heart rate in the healthy range (i.e., 40-80~bpm) does not produce significant changes in the flow dynamics. The values of heart rates employed in this work are relatively small but still represent a physiological range at rest. Simulations could be carried out with higher heart rate values, but this requires a more complex elastic model, together with a higher space and temporal resolution.

Finally, we explore the effect of aortic valve stiffness by comparing a physiologically compliant valve against a stiffened valve, representative of stenotic conditions. The stenotic valve leads to a significant increase in velocity intermittency in both the aorta and ventricle, as indicated by elevated flatness levels. This behavior is likely driven by the narrower valve opening and the associated sharp velocity gradients, which intensify local flow fluctuations. 

Our findings demonstrate that Lagrangian statistics offer a powerful and sensitive framework for probing the complex dynamics of cardiovascular flow. By systematically varying physiological parameters such as heart rate and valve stiffness, we have shown how key features of turbulence and intermittency are modulated within the heart. These insights may prove valuable for future clinical investigations of pathological flow conditions and for guiding the design of prosthetic valves~\cite{scarpolini2025hemodynamiceffectsintrasupra}. 
Moreover, the high resolution and detailed velocity statistics accessible through our Lagrangian framework open the possibility of simulating and studying the dynamics of individual RBCs at the cellular scale~\cite{guglietta2020effects}, enabling more accurate predictions of phenomena such as hemolysis and targeted drug delivery.

\subsection*{Acknowledgments}
This work was supported by the Italian Ministry of University and Research (MUR) under the FARE programme (No. R2045J8XAW), project “Smart-HEART”, and by the European Research Council (ERC) under the European Union’s Horizon 2020 research and innovation program (Grant Agreement No. 882340).
Luca Biferale and Fabio Guglietta acknowledge funding from Tor Vergata University project AI4HEART. 
Francesco Viola acknowledges financial support by the European Research Council (ERC) under the European Union’s Horizon Europe research and innovation program, Project CARDIOTRIALS (Grant Agreement No. 101039657).

\printbibliography
\end{document}